\documentclass[12pt]{article}
\usepackage{amsmath}
\usepackage{graphicx}
\usepackage{ulem}

\newcommand{\be}{\begin{eqnarray}}
\newcommand{\ee}{\end{eqnarray}}

\begin{document}

\title{\textbf{Azimuthal correlations from transverse momentum conservation
and possible local parity violation}}
\author{Adam Bzdak$^{a,b}$, Volker Koch$^{a}$, Jinfeng Liao$^{a}$ \\
\\
$^{a}$ Lawrence Berkeley National Laboratory, 1 Cyclotron Road\\
MS70R0319, Berkeley, CA 94720, USA\thanks{%
e-mails: ABzdak@lbl.gov, VKoch@lbl.gov, JLiao@lbl.gov} \\
$^{b}$ Institute of Nuclear Physics, Polish Academy of Sciences\\
Radzikowskiego 152, 31-342 Krakow, Poland }
\maketitle

\begin{abstract}
We analytically calculate the contribution of transverse momentum
conservation to the azimuthal correlations that have been proposed as
signals for possible local strong parity violation and recently measured in
heavy ion collisions. These corrections are of the order of the inverse of
the total final state particle multiplicity and thus are of the same order
as the observed signal. The corrections contribute with the same sign to
both like-sign and opposite-sign pair correlations. Their dependence on the
momentum is in qualitative agreement with the measurements by the STAR
collaboration, while the pseudorapidity dependence differs from the data.

\vskip 0.5cm

\end{abstract}

\newpage

\section{Introduction}

Topological configurations occur generically in non-Abelian gauge theories
and are known to be essential for understanding the vacuum structure and
hadron properties in Quantum Chromodynamics (QCD) (for reviews see e.g. \cite%
{topology_vacuum}). They have also been shown to play important roles in hot
QCD matter (i.e. quark-gluon plasma) existing in the early universe and now
created in heavy ion collisions \cite{topology_hot}. Despite many indirect
evidences, a direct experimental manifestation of the topological effects
has not been achieved and is therefore of great interest. One salient
feature of the topological configurations is the $\mathcal{P}$- and $%
\mathcal{CP}$-odd effects they may induce. Based on that, it has been
suggested \cite{lpv,cme1,cme2,lpv-rev} to look for possible occurrence of $%
\mathcal{P}$- and $\mathcal{CP}$-odd domains with local strong parity
violation for a direct detection of topological effects. Such domains may
naturally arise in a heavy ion collision due to the so-called sphaleron
transitions in the created hot QCD matter. In particular, the so called
Chiral Magnetic Effect (CME) predicts that in the presence of the strong
external (electrodynamic) magnetic field at the early stage after a
(non-central) collision, sphaleron transitions induce a separation of
negatively and positively charged particles along the direction of the
magnetic field which is perpendicular to the reaction plane defined by the
impact parameter and the beam axis. Such an out-of-plane charge separation,
however, varies its orientation from event to event, either parallel or
anti-parallel to the magnetic field (depending whether the CME is caused by
sphaleron or antisphaleron transition). As a result the expectation value of
any $\mathcal{P}$-odd observable vanishes and only the variance of such
observable may be detected, making the measurement of CME rather
challenging. Recently the STAR collaboration has published in \cite{star-p}
measurements of charged particle azimuthal correlations proposed in \cite{SV}
as a CME signal, and found interesting patterns partly consistent with CME
expectations. The STAR data have generated considerable interests and many
subsequent works have appeared, proposing alternative explanations \cite%
{FW,SS-SP,SP,Asakawa:2010bu}, suggesting data interpretations and new
observations \cite{MS,BKL,LKB,Voloshin:2010ut}, and studying further
consequences of CME \cite{spiral,suscep,Kang:2010qx}.

We start with a discussion of the proposed CME signal as measured by STAR.
In Ref. \cite{SV} it has been suggested that the CME may be indirectly
approached by the measurement of the following two-particle correlation 
\begin{eqnarray}
\gamma &=&\left\langle \cos (\phi _{1}+\phi _{2}-2\Psi _{RP})\right\rangle 
\notag \\
&=&\left\langle \cos (\phi _{1}-\Psi _{RP})\cos (\phi _{2}-\Psi
_{RP})\right\rangle -\left\langle \sin (\phi _{1}-\Psi _{RP})\sin (\phi
_{2}-\Psi _{RP})\right\rangle ,  \label{gam}
\end{eqnarray}%
where $\Psi _{RP},$ $\phi _{1}$ and $\phi _{2}$ denote the azimuthal angles
of the reaction plane and produced charged particles, respectively. Since
this observable measures the difference between the in-plane and
out-of-plane projected azimuthal correlations, it has been argued that this
observable is particularly suited for revealing the CME signal which is an
out-of-plane charge separation. Specifically, the CME predicts $\gamma >0$
for opposite sign-pairs and $\gamma <0$ for same-sign pairs. However, as has
been pointed out in \cite{FW,SS-SP,SP,BKL,LKB,Voloshin:2010ut}, in
non-central collisions the presence of elliptic flow \cite{Voloshin:2008dg}
already differentiates between in- and out-of-plane. As a consequence,
essentially any two-particle correlation will contribute to the above
observable even though the correlations' dynamical mechanism may generally
be reaction-plane independent. These so-called ``background'' correlations,
need to be well understood theoretically and possibly be determined by
independent measurements.\footnote{%
It is worth emphasizing that the contributions to $\gamma $ from elliptic
flow induced correlations and from the CME have similar centrality trends:
both the elliptic flow and the magnetic field, necessary for the CME,
increase from central to peripheral collisions.} The STAR publication \cite%
{star-p} presents data for the correlator $\gamma $ together with the
reaction plane independent correlator 
\begin{equation}
\delta =\left\langle \cos \left( \phi _{1}-\phi _{2}\right) \right\rangle
\label{delta}
\end{equation}
in the midrapidity region for both same- and opposite-sign pairs in $AuAu$
and $CuCu$ collisions at two energies $\sqrt{s_{NN}}=200$ and $62$ GeV. The
data are encouraging: at first sight the results for $\gamma $ seem to be
qualitatively consistent with the CME expectations. However, as shown in %
\cite{BKL} when taking the correlator $\delta $ into account as well, the
interpretation of the data in term of the CME requires almost exact
cancellation of the CME and all possible ``background'' correlations.
Consequently in order to extract a possible signal for the CME, the
understanding of these ``background'' correlations becomes crucial at this
stage.

One well-known possible source of azimuthal correlation is the conservation
of transverse momentum, which has been qualitatively discussed in \cite{SP}
and has been suggested to be a significant contribution to the measured
observable $\gamma $. The argument goes as follows. Consider for a moment 
\textit{all} particles in the final state, charged and neutral over all
phase-space. Next rewrite the correlator (\ref{gam}) as (for simplicity we
set $\Psi _{RP}=0$ here and in the rest of the paper) 
\begin{align}
\gamma & =\left\langle \frac{\sum\nolimits_{i\neq j}\cos (\phi _{i}+\phi
_{j})}{\sum_{i\neq j}1}\right\rangle  \notag \\
& =\left\langle \frac{\left( \sum\nolimits_{i}\cos (\phi _{i})\right)
^{2}-\left( \sum\nolimits_{i}\sin (\phi _{i})\right)
^{2}-\sum\nolimits_{i}\cos (2\phi _{i})}{\sum_{i\neq j}1}\right\rangle ,
\label{gam_dif}
\end{align}%
where $i$ and $j$ are summed over all particles. If we further assume that
all particles have exactly the same magnitude of transverse momentum $p_{t}$%
, the conservation of transverse momentum implies 
\begin{equation}
\sum\nolimits_{i}\cos (\phi _{i})=\sum\nolimits_{i}\sin (\phi _{i})=0,
\end{equation}%
and in consequence one obtains for sufficiently large $N$ 
\begin{equation}
\gamma =\frac{-v_{2}}{N}.  \label{gam_int}
\end{equation}%
Here $v_{2}$ is the elliptic flow coefficient measured for all produced
particles and $N$ is the total number of all produced particles (in full
phase space). This contribution to the azimuthal correlations from
transverse momentum conservation turns out to be at the order of the data
measured by STAR, and therefore bears interest and importance.

However, the above argument relies on two assumptions which are not realized
in the actual measurement. First, STAR measures only the charged particles
in a small pseudorapidity region $|\eta |<1$, accounting only for about $%
15\% $ of the total number of produced particles. Second, the magnitude of
the transverse momentum is not a constant but rather distributed, more or
less according to a thermal distribution. Therefore, a more realistic
estimate for the contribution of transverse momentum conservation to the
above correlation functions is required. The influence of momentum
conservation on observables in heavy ion collisions has been discussed in
the literature in the context of spectra \cite{CL}, elliptic flow \cite%
{BDO,CL} and certain two-particle densities \cite{Trai}, and corrections of
various importance have been established. In the present paper, we will
address the effect of transverse momentum conservation on the correlation
functions relevant for the potential measurement of the CME. To this end we
derive the necessary formalism which allows us to quantify the azimuthal
correlations due to transverse momentum conservation.

Before going into details, let us discuss a few qualitative features which
are to be expected from transverse momentum conservation, and which will be
demonstrated in detail below. First, transverse momentum conservation
introduces a back-to-back correlation for particle pairs, as they tend to
balance each other in momentum. Second, the expected correction should scale
inversely with total number of particles, as more particles provide more
ways to balance the momentum and thus dilute the effect on two-particle
correlations. Furthermore the correlation should be stronger in-plane than
out-of-plane due to the presence of elliptic flow. As a result we expect
that transverse momentum conservation results in a negative contribution to
the observable $\gamma $, which increase with the strength of the elliptic
flow, $v_{2}$. Finally, we note that transverse momentum conservation is
``blind'' to particle charge, leading to identical contributions to
same-sign and opposite-sign pair-correlations. Because of these features,
transverse momentum conservation \textit{alone }cannot be expected as a full
account for the observed charged particle azimuthal correlation patterns. It
should be rather considered as an important background effect that
contributes significantly and, therefore, necessitates quantitative studies
for establishing any final interpretation of the data.

The paper is organized as follows. In the next Section, assuming only
transverse momentum conservation, we present analytical calculations of the
correlators $\gamma $ and $\delta $. In Section $3$ we take into account the
STAR acceptance and compare our results with the available data (integrated
over transverse momentum). In Section $4$ we discuss specific differential
observables, such as transverse momentum and rapidity dependent
correlations. Our conclusions are listed in the last section, where also
some comments are included.

\section{General formulas}

Let us assume that there are $N$ particles in total produced in a given
heavy ion collision event, with individual momenta $\vec{p}_{1},...,\vec{p}%
_{N}$. We denote the particles' transverse momenta by $\vec{p}_{1,t},...,%
\vec{p}_{N,t}$ and their magnitudes by $p_{1,t},...,p_{N,t}$. The $N$
particle density $f_{N}$ (normalized to unity) with enforced transverse
momentum conservation can be written as 
\begin{equation}
f_{N}(\vec{p}_{1},...,\vec{p}_{N})=\frac{\delta ^{2}\left( \vec{p}_{1,t}+...+%
\vec{p}_{N,t}\right) f(\vec{p}_{1})...f(\vec{p}_{N})}{\int_{F}\delta
^{2}\left( \vec{p}_{1,t}+...+\vec{p}_{N,t}\right) f(\vec{p}_{1})...f(\vec{p}%
_{N})d^{3}\vec{p}_{1}...d^{3}\vec{p}_{N}},  \label{f_N}
\end{equation}%
where $f(\vec{p}_{i})$ is the normalized ($\int_{F}f(\vec{p})d^{3}\vec{p}=1$%
) single particle distribution. Note that the above integrals are taken over
the full phase space (denoted by ``$F$'') rather than the region where
particles are actually measured. In Eq. (\ref{f_N}) we explicitly assume
that all produced particles are governed by the same single particle
distribution \footnote{%
This assumption is reasonably well satisfied in heavy ion collisions, where
the final state particles are mostly pions.}. We also ignore any other
two-particle correlations, since in this paper we focus on the effects of
transverse momentum conservation only. To calculate a two-particle
correlator, such as $\left\langle \cos (\phi _{1}+\phi _{2})\right\rangle $,
we need the two-particle density which can be obtained from Eq. (\ref{f_N})
by integrating out $N-2$ momenta%
\begin{equation}
f_{2}\left( \vec{p}_{1},\vec{p}_{2}\right) =f(\vec{p}_{1})f(\vec{p}_{2})%
\frac{\int_{F}\delta ^{2}\left( \sum\nolimits_{i=1}^{N}\vec{p}_{i,t}\right)
\prod\nolimits_{i=3}^{N}\left[ f(\vec{p}_{i})d^{3}\vec{p}_{i}\right] }{%
\int_{F}\delta ^{2}\left( \sum\nolimits_{i=1}^{N}\vec{p}_{i,t}\right)
\prod\nolimits_{i=1}^{N}\left[ f(\vec{p}_{i})d^{3}\vec{p}_{i}\right] }.
\label{f_2}
\end{equation}

In order to perform the integrals in Eq. (\ref{f_2}) we follow the
techniques of Refs. \cite{CL,BDO,D} and make use of the central limit
theorem. The sum of $M$ uncorrelated transverse momenta $\sum_{i=1}^{M}\vec{p%
}_{i,t}=\vec{K}_{t}$ has a Gaussian distribution if $M$ is sufficiently large%
\footnote{%
We have checked by simple MC calculations that the central limit theorem
applies for $M>10$, see also \cite{CL}. This condition is very well
satisfied in heavy ion collisions, even in the most peripheral ones.}, that
is 
\begin{align}
G_{M}(\vec{K}_{t})& =\int_{F}\delta ^{2}\left( \sum\nolimits_{i=1}^{M}\vec{p}%
_{i,t}-\vec{K}_{t}\right) \prod\nolimits_{i=1}^{M}\left[ f(\vec{p}_{i})d^{3}%
\vec{p}_{i}\right]   \notag \\
& =\frac{1}{2\pi M\sqrt{\left\langle p_{x}^{2}\right\rangle _{F}\left\langle
p_{y}^{2}\right\rangle _{F}}}\exp \left( -\frac{K_{x}^{2}}{2M\left\langle
p_{x}^{2}\right\rangle _{F}}-\frac{K_{y}^{2}}{2M\left\langle
p_{y}^{2}\right\rangle _{F}}\right) .  \label{G_M}
\end{align}%
Here $x$ and $y$ denote the two components of transverse momentum and 
\begin{equation}
\left\langle p_{x}^{2}\right\rangle _{F}=\int_{F}f(\vec{p})p_{x}^{2}d^{3}%
\vec{p},\quad \left\langle p_{y}^{2}\right\rangle _{F}=\int_{F}f(\vec{p}%
)p_{y}^{2}d^{3}\vec{p},
\end{equation}%
where the integrations are over full phase space $F$. Using Eq. (\ref{G_M})
we can express Eq. (\ref{f_2}) in the following way%
\begin{align}
f_{2}(\vec{p}_{1},\vec{p}_{2})& =f(\vec{p}_{1})f(\vec{p}_{2})\frac{%
G_{N-2}\left( -\vec{p}_{1,t}-\vec{p}_{2,t}\right) }{G_{N}(0)}  \notag \\
& =f(\vec{p}_{1})f(\vec{p}_{2})\frac{N}{N-2}\exp \left( -\frac{%
(p_{1,x}+p_{2,x})^{2}}{2(N-2)\left\langle p_{x}^{2}\right\rangle _{F}}-\frac{%
(p_{1,y}+p_{2,y})^{2}}{2(N-2)\left\langle p_{y}^{2}\right\rangle _{F}}%
\right) .  \label{f_2exp}
\end{align}%
Expanding in powers of $1/N$ and restricting ourselves to pairs with $\left| 
\vec{p}_{1,t}+\vec{p}_{2,t}\right| \ll \sqrt{2N\left\langle
p_{t}^{2}\right\rangle }$ \footnote{%
In practice this is hardly any restriction, since for sufficiently large $N$
only pairs with very large transverse momentum violate the condition $\left| 
\vec{p}_{1,t}+\vec{p}_{2,t}\right| \ll \sqrt{2N\left\langle
p_{t}^{2}\right\rangle }$, which are strongly suppressed by the
exponentially decreasing single particle distributions, $f(\vec{p}_{1})f(%
\vec{p}_{2})$.} we obtain 
\begin{equation}
f_{2}(\vec{p}_{1},\vec{p}_{2})\simeq f(\vec{p}_{1})f(\vec{p}_{2})\left( 1+%
\frac{2}{N}-\frac{(p_{1,x}+p_{2,x})^{2}}{2N\left\langle
p_{x}^{2}\right\rangle _{F}}-\frac{(p_{1,y}+p_{2,y})^{2}}{2N\left\langle
p_{y}^{2}\right\rangle _{F}}\right) .  \label{f_2end}
\end{equation}%
As already mentioned, the above correlation function does not distinguish
between same- and opposite-sign pairs (or pairs involving neutral
particles), as the transverse momentum conservation involves all particles
without discrimination on the charge of particles. We also note that the
two-particle density, $f_{2}(\vec{p}_{1},\vec{p}_{2})$ is maximum for
back-to-back configurations, as can be easily seen from the dependence on
the combinations $-(p_{1,x}+p_{2,x})^{2}$ and $-(p_{1,y}+p_{2,y})^{2}$.

Given the two-particle density Eq. (\ref{f_2},\ref{f_2end}) we can proceed
to evaluate various two-particle azimuthal correlations in any given
kinematic region, for example the one introduced in Eq. (\ref{gam}) 
\begin{equation}
\left\langle \cos (\phi _{1}+\phi _{2})\right\rangle =\frac{%
\int\limits_{\Omega }f_{2}(\vec{p}_{1},\vec{p}_{2})\cos (\phi _{1}+\phi
_{2})d^{3}\vec{p}_{1}d^{3}\vec{p}_{2}}{\int\limits_{\Omega }f_{2}(\vec{p}%
_{1},\vec{p}_{2})d^{3}\vec{p}_{1}d^{3}\vec{p}_{2}},
\end{equation}%
and analogously for $\left\langle \cos (\phi _{1}-\phi _{2})\right\rangle $.
Here we denote by $\Omega $ the part of the phase space covered by the
actual experiment.

Finally we need the single particle distribution, for which we assume the
following rather general form ($\Psi _{RP}=0$): 
\begin{equation}
f\left( \vec{p}\right) d^{3}\vec{p}=\frac{g(p_{t},\eta )}{2\pi }\left[
1+2v_{2}(p_{t},\eta )\cos (2\phi )\right] d^{2}\vec{p}_{t}d\eta ,
\end{equation}%
where $v_{2}$ is the $p_{t}$ and $\eta $ dependent elliptic flow coefficient
(with $\eta $ the pseudorapidity). Taking Eq. (\ref{f_2end}) into account
and performing elementary calculations we obtain our main result \footnote{%
Here and in the following we assume that azimuthal part of $\Omega $ covers
full $2\pi $ i.e., $\phi \in \lbrack 0,2\pi )$. This is the only restriction
we impose on $\Omega $.}:%
\begin{equation}
\left\langle \cos (\phi _{1}+\phi _{2})\right\rangle =-\frac{1}{N}\frac{%
\left\langle p_{t}\right\rangle _{\Omega }^{2}}{\left\langle
p_{t}^{2}\right\rangle _{F}}\frac{2\bar{v}_{2,\Omega }-\bar{\bar{v}}_{2,F}-%
\bar{\bar{v}}_{2,F}(\bar{v}_{2,\Omega })^{2}}{1-\left( \bar{\bar{v}}%
_{2,F}\right) ^{2}},  \label{main}
\end{equation}%
where we have introduced certain weighted moments of $v_{2}$ 
\begin{equation}
\bar{v}_{2}=\frac{\left\langle v_{2}(p_{t},\eta )p_{t}\right\rangle }{%
\left\langle p_{t}\right\rangle }=\frac{\int g(p_{t},\eta )v_{2}(p_{t},\eta
)p_{t}d^{2}\vec{p}_{t}d\eta }{\int g(p_{t},\eta )p_{t}d^{2}\vec{p}_{t}d\eta }%
,  \label{v2-}
\end{equation}%
and%
\begin{equation}
\bar{\bar{v}}_{2}=\frac{\left\langle v_{2}(p_{t},\eta
)p_{t}^{2}\right\rangle }{\left\langle p_{t}^{2}\right\rangle }=\frac{\int
g(p_{t},\eta )v_{2}(p_{t},\eta )p_{t}^{2}d^{2}\vec{p}_{t}d\eta }{\int
g(p_{t},\eta )p_{t}^{2}d^{2}\vec{p}_{t}d\eta }.  \label{v2=}
\end{equation}%
Again, the indexes $F$ and $\Omega $ indicate that all integrations in Eqs. (%
\ref{v2-}), (\ref{v2=}) are performed over full phase space ($F$) or the
phase space where particles are measured ($\Omega $), respectively. For
completeness let us add that $N$ denotes the total number of produced
particles (charged and neutral) and $\left\langle p_{t}\right\rangle
_{\Omega }$ is the average transverse momentum of the \textit{measured}
particles.

One important lesson from the above result is that even if we measure
particles in a limited fraction of the full phase space, e.g. a narrow
pseudorapidity bin, the effect of transverse momentum conservation on $%
\left\langle \cos (\phi _{1}+\phi _{2})\right\rangle $ is \textit{not}
suppressed\footnote{%
This statement is of course limited by the applicability of the central
limit theorem, which however works for as little as $10$ particles.}. This
may be roughly understood in the following way: if each particle is
generated in an independent manner except the constraint from overall
transverse momentum conservation, then for each given particle its $p_{t}$
effectively has a chance of $1/N$ (in the large $N$ limit) to be balanced by
every other particle. That is equivalent to say that each pair has a
back-to-back correlation of strength $1/N$, which is preserved despite what
fraction of particles is selected for measurement. Furthermore, while not
changing the order of magnitude of the effect, the details of the $p_{t}$
and $\eta $ dependence of single particle distribution and $v_{2}$ may
slightly affect the quantitative results.

Next we examine two approximations to the main result, Eq. (\ref{main}).

\begin{itemize}
\item[(i)] If all the produced particles are measured, i.e. $\Omega =F$, and
all have the same magnitude of the transverse momentum i.e. $g(p_{t},\eta
)\propto \delta (p_{t}-p_{0})h(\eta )$, we recover the result (\ref{gam_int}%
) discussed in the Introduction.

\item[(ii)] If one allows for a finite acceptance but neglects the $p_{t}$
and $\eta $ dependence of $v_{2}$, i.e. $\bar{\bar{v}}_{2,F/\Omega }=\bar{v}%
_{2,F/\Omega }=v_{2,F/\Omega }$, Eq. (\ref{main}) reduces to 
\begin{equation}
\left\langle \cos (\phi _{1}+\phi _{2})\right\rangle =-\frac{v_{2}}{N}\frac{%
\left\langle p_{t}\right\rangle _{\Omega }^{2}}{\left\langle
p_{t}^{2}\right\rangle _{F}}.
\end{equation}%
Since $\frac{\left\langle p_{t}\right\rangle _{\Omega }^{2}}{\left\langle
p_{t}^{2}\right\rangle _{F}}$ depends only weakly on the acceptance, $\Omega 
$, the corrections due to transverse momentum conservation are, as already
pointed out, more or less independent of the number of observed particles.
\end{itemize}

Next let us calculate the contribution from transverse momentum conservation
to the correlation function $\delta =\left\langle \cos (\phi _{1}-\phi
_{2})\right\rangle $ which has also been measured by STAR. Following similar
procedures, we obtain the following result: 
\begin{equation}
\left\langle \cos (\phi _{1}-\phi _{2})\right\rangle =-\frac{1}{N}\,\frac{%
\left\langle p_{t}\right\rangle _{\Omega }^{2}}{\left\langle
p_{t}^{2}\right\rangle _{F}}\,\frac{1+(\bar{v}_{2,\Omega })^{2}-2\bar{\bar{v}%
}_{2,F}\,\bar{v}_{2,\Omega }}{1-\left( \bar{\bar{v}}_{2,F}\right) ^{2}},
\label{eq_minus}
\end{equation}%
We notice a few interesting features. First, the correlation scales like $%
-1/N$ as expected. Second, the effect does not depend on elliptic flow in
leading order, and thus, is much stronger than that in $\left\langle \cos
(\phi _{1}+\phi _{2})\right\rangle $ which is of order $\hat{O}(v_{2}/N)$.
Let us again examine the same two approximations to the above result.

\begin{itemize}
\item[(i)] If all the produced particles are measured, i.e. $\Omega =F$, and
all have the same magnitude of transverse momentum i.e. $g(p_{t},\eta
)\propto \delta (p_{t}-p_{0})h(\eta )$, we obtain $\left\langle \cos (\phi
_{1}-\phi _{2})\right\rangle =-\frac{1}{N}$.

\item[(ii)] If one allows for a finite acceptance but neglects the $p_{t}$
and $\eta $ dependence of $v_{2}$, i.e. $\bar{\bar{v}}_{2,F/\Omega }=\bar{v}%
_{2,F/\Omega }=v_{2,F/\Omega }$, we obtain%
\begin{equation}
\left\langle \cos (\phi _{1}-\phi _{2})\right\rangle =-\frac{1}{N}\frac{%
\left\langle p_{t}\right\rangle _{\Omega }^{2}}{\left\langle
p_{t}^{2}\right\rangle _{F}}.
\end{equation}
\end{itemize}

Thus the main difference between the two correlators, $\gamma$ (Eq. \ref%
{main}) and $\delta$ (Eq. \ref{eq_minus}) is the presence of $v_2$ in the
former. Therefore, effects of transverse momentum conservation will be much
more visible in the correlator $\left\langle \cos (\phi _{1}-\phi _{2})
\right\rangle $.

\section{Comparison with data}

In this Section we compare our results (\ref{main}) and (\ref{eq_minus})
with recently published STAR data \cite{star-p}, where charged particles
have been measured in the pseudorapidity interval $-1<\eta <1$ and the
transverse momentum region $p_{t}>0.15$ GeV. For the results integrated over
transverse momentum additional cut was imposed $p_{t}<2$ GeV. As seen from
Eqs. (\ref{main},\ref{v2-},\ref{v2=},\ref{eq_minus}) to calculate the
contribution of the transverse momentum conservation to $\left\langle \cos
(\phi _{1}+\phi _{2})\right\rangle $ and $\left\langle \cos (\phi _{1}-\phi
_{2})\right\rangle $ we need full information about single particle
distribution $g(p_{t},\eta )$ and elliptic flow $v_{2}(p_{t},\eta )$ in the
full phase space. Unfortunately such complete information is not currently
available. Thus we will make some reasonable assumptions that hopefully
allow us to obtain an approximate insight into the discussed effect.

First let us estimate the total number of produced particles $N$. From the
PHOBOS measurement \cite{Ntot} we know that the total number of charged
particles $N_{ch}$ grows linearly\footnote{%
In contrast to the number of charged particles at midrapidity that grows
slightly faster than $N_{part}$.} with the number of participants $N_{part}$
(or equivalently number of wounded nucleons \cite{BBC}). At $\sqrt{s_{NN}}%
=200$ GeV we find that $N_{ch}\approx 14N_{part}$ \cite{Ntot}, thus the
total number of particles can be reasonably approximated by 
\begin{equation}
N\approx (3/2)N_{ch}=21N_{part}.
\end{equation}

This result allows us to roughly estimate the contribution of transverse
momentum conservation. We simply assume that $\left\langle
p_{t}\right\rangle _{\Omega }^{2}=\left\langle p_{t}^{2}\right\rangle _{F}$
and $\bar{v}_{2,\Omega }=\bar{\bar{v}}_{2,F}$ which leads to 
\begin{equation}
\left\langle \cos (\phi _{1}+\phi _{2})\right\rangle \cdot N_{part}=-\bar{v}%
_{2,\Omega }/21\approx -0.004,  \label{res_p}
\end{equation}%
and%
\begin{equation}
\left\langle \cos (\phi _{1}-\phi _{2})\right\rangle \cdot
N_{part}=1/21\approx -0.05,  \label{res_m}
\end{equation}%
where we take $\bar{v}_{2,\Omega }=0.08$ to be slightly larger than the
elliptic flow parameter $v_{2,\Omega }=0.06$ in order to account for the
momentum dependence of $v_{2}$. Below we will show that these simple
assumptions are well reproduced in a more detailed calculation.

While this effect gives a contribution with the same sign and
order-of-magnitude for the same-sign pair correlation data, it is a factor $%
3-5$ (very peripheral -- mid-central) less in magnitude for $\left\langle
\cos (\phi _{1}+\phi _{2})\right\rangle $ and a factor $1.5-4$ (mid-central
-- very peripheral) larger than the STAR data. It also gives the same
contribution to the opposite-sign pair correlation for which the data are
positive.

To perform more precise calculations we further assume that the single
particle distribution $g(p_{t},\eta )$ can be expressed in the following way%
\begin{equation}
g(p_{t},\eta )=\frac{1}{T^{2}}\exp \left( \frac{-p_{t}}{T}\right) h(\eta ),
\label{g}
\end{equation}%
where for simplicity we take a thermal distribution and assume factorization
of the momentum and pseudorapidity dependence. To introduce a slight
pseudorapidity dependence of $T$ \cite{brahms} we will distinguish between $%
T_{F}$ and $T_{mid}$ depending upon whether we integrate over full
pseudorapidity range or the midrapidity region. In the following we take $%
T_{mid}=0.225$ GeV and $T_{mid}/T_{F}=1.1$ \cite{brahms}. In Eq. (\ref{g}) $%
h(\eta )$ is the normalized pseudorapidity single particle distribution,
which can be well represented by a double Gaussian form 
\begin{equation}
h(\eta )=\frac{1}{2\sqrt{2\pi \sigma ^{2}}}\left[ \exp \left( -\frac{(\eta
-\eta _{0})^{2}}{2\sigma ^{2}}\right) +\exp \left( -\frac{(\eta +\eta
_{0})^{2}}{2\sigma ^{2}}\right) \right] ,  \label{h}
\end{equation}%
with $\eta _{0}=2$ and $\sigma =1.9$ for $\sqrt{s_{NN}}=200$ GeV.\footnote{%
We performed the fit to the PHOBOS $35-45\%$ centrality $AuAu$ data \cite%
{PHO-fit}.} Finally we have to specify the momentum and pseudorapidity
dependence of elliptic flow. For simplicity we represent this dependence
with a factorized liner ansatz for both the transverse momentum and the
pseudorapidity, which represents the presently available date reasonably
well up to $p_{t}\simeq 2$ GeV \cite{v2-PHO,v2-STAR,v2-PHEN} \footnote{%
We have checked the influence of constant $v_{2}$ for $p_{t}>2$ GeV and
found it negligible.} 
\begin{equation}
v_{2}(p_{t},\eta )=Cp_{t}\frac{7-|\eta |}{7},  \label{v2_app}
\end{equation}%
and $v_{2}=0$ for $|\eta |>7$. Here $C=0.14$ so that for the midrapidity
region $|\eta |<1$ the calculated elliptic flow equals $0.06$ \cite%
{v2-PHO,v2-STAR,v2-PHEN}.

With these parametrisations we find that: $\left\langle p_{t}\right\rangle
_{\Omega }^{2}\approx 0.257$, $\left\langle p_{t}^{2}\right\rangle
_{F}\approx 0.252$, $\bar{v}_{2,\Omega }\approx 0.0886$ and $\bar{\bar{v}}%
_{2,F}\approx 0.0773$. Substituting these number into Eqs. (\ref{main}), (%
\ref{eq_minus}) we obtain%
\begin{equation}
\left\langle \cos (\phi _{1}+\phi _{2})\right\rangle \cdot N_{part}\approx
-0.005,
\end{equation}%
and%
\begin{equation}
\left\langle \cos (\phi _{1}-\phi _{2})\right\rangle \cdot N_{part}\approx
-0.05,
\end{equation}%
which are very close to the numbers estimated at the beginning of this
Section. While it is difficult to estimate the precise uncertainty of our
calculation, however, we expect our results to be correct within a few tens
of percent.

\section{Differential distributions}

The STAR collaboration also presented data for $\left\langle \cos (\phi
_{1}+\phi _{2})\right\rangle $ as a function of $p_{+}=(p_{1,t}+p_{2,t})/2$
and $p_{-}=\left| p_{1,t}-p_{2,t}\right| $. Both distributions are very
informative: the azimuthal correlation $\left\langle \cos (\phi _{1}+\phi
_{2})\right\rangle $ increases roughly linearly with $p_{+}$ while it
depends only weakly on $p_{-}$. In \cite{BKL} we showed that such behavior
is not inconsistent with the CME and can be understood if we assume that
correlated pairs have slightly larger momenta than uncorrelated ones.

As we will show below, a qualitatively similar behaviour can be obtained
from transverse momentum conservation. With the two-particle distribution (%
\ref{f_2end}), the $p_{+}$ differential distribution reads%
\begin{equation}
\left\langle \cos (\phi _{1}+\phi _{2})\right\rangle _{p_{+}}=\frac{%
\int\limits_{\Omega }f_{2}(\vec{p}_{1},\vec{p}_{2})\delta \left(
2p_{+}-p_{1,t}-p_{2,t}\right) \cos (\phi _{1}+\phi _{2})d^{3}\vec{p}_{1}d^{3}%
\vec{p}_{2}}{\int\limits_{\Omega }f_{2}(\vec{p}_{1},\vec{p}_{2})\delta
\left( 2p_{+}-p_{1,t}-p_{2,t}\right) d^{3}\vec{p}_{1}d^{3}\vec{p}_{2}}
\label{pt-dep}
\end{equation}%
and analogously for $p_{-}=\left| p_{1,t}-p_{2,t}\right| $.

Let us first discus the simplified case where $v_{2}(p_{t},\eta )$ is
replaced by its average value $v_{2}$. Taking Eqs. (\ref{f_2end}) and (\ref%
{g}) into account we obtain\footnote{%
In this section for simplicity we integrate from $p_{t}=0$ despite the
finite cut $p_{t}>0.15$ GeV in the STAR experiments.}%
\begin{equation}
\left\langle \cos (\phi _{1}+\phi _{2})\right\rangle _{p_{+}}=-v_{2}\frac{%
2p_{+}^{2}}{15NT^{2}},
\end{equation}%
and 
\begin{equation}
\left\langle \cos (\phi _{1}+\phi _{2})\right\rangle _{p_{-}}=-v_{2}\frac{%
3T^{2}+3Tp_{-}+p_{-}^{2}}{6NT\left( T+p_{-}\right) },
\end{equation}%
where for simplicity we set $T_{F}=T_{mid}=T$. As can be seen, the
correlation shows a strong quadratic growth with increasing $p_{+}$, in
qualitative agreement with data. The dependence on $p_{-}$, on the other
hand, is essentially constant for $p_{-}\ll T$ before it exhibits a linear
increase. Performing analogous calculations for $\delta $ we obtain%
\begin{equation}
\left\langle \cos (\phi _{1}-\phi _{2})\right\rangle _{p_{+}/p_{-}}=\frac{1}{%
v_{2}}\left\langle \cos (\phi _{1}+\phi _{2})\right\rangle _{p_{+}/p_{-}},
\end{equation}%
thus the $p_{+}$ and $p_{-}$ dependence is identical however, the signal is
significantly stronger.

\begin{figure}[h]
\begin{center}
\includegraphics[scale=0.8]{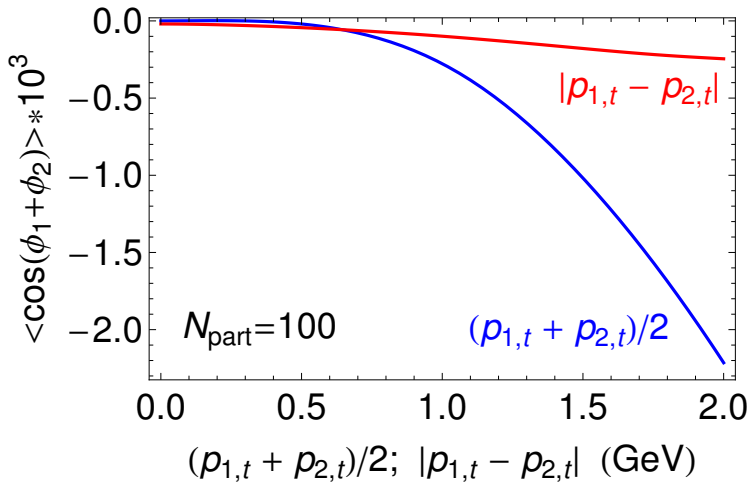}
\end{center}
\caption{(Color online) The two-particle azimuthal correlation $\left\langle
\cos (\protect\phi _{1}+\protect\phi _{2})\right\rangle $ vs $%
p_{+}=(p_{1,t}+p_{2,t})/2$ (blue line) and $p_{-}=\left|
p_{1,t}-p_{2,t}\right| $ (red line) for $N_{part}=100$.}
\label{fig1}
\end{figure}

\begin{figure}[h]
\begin{center}
\includegraphics[scale=0.8]{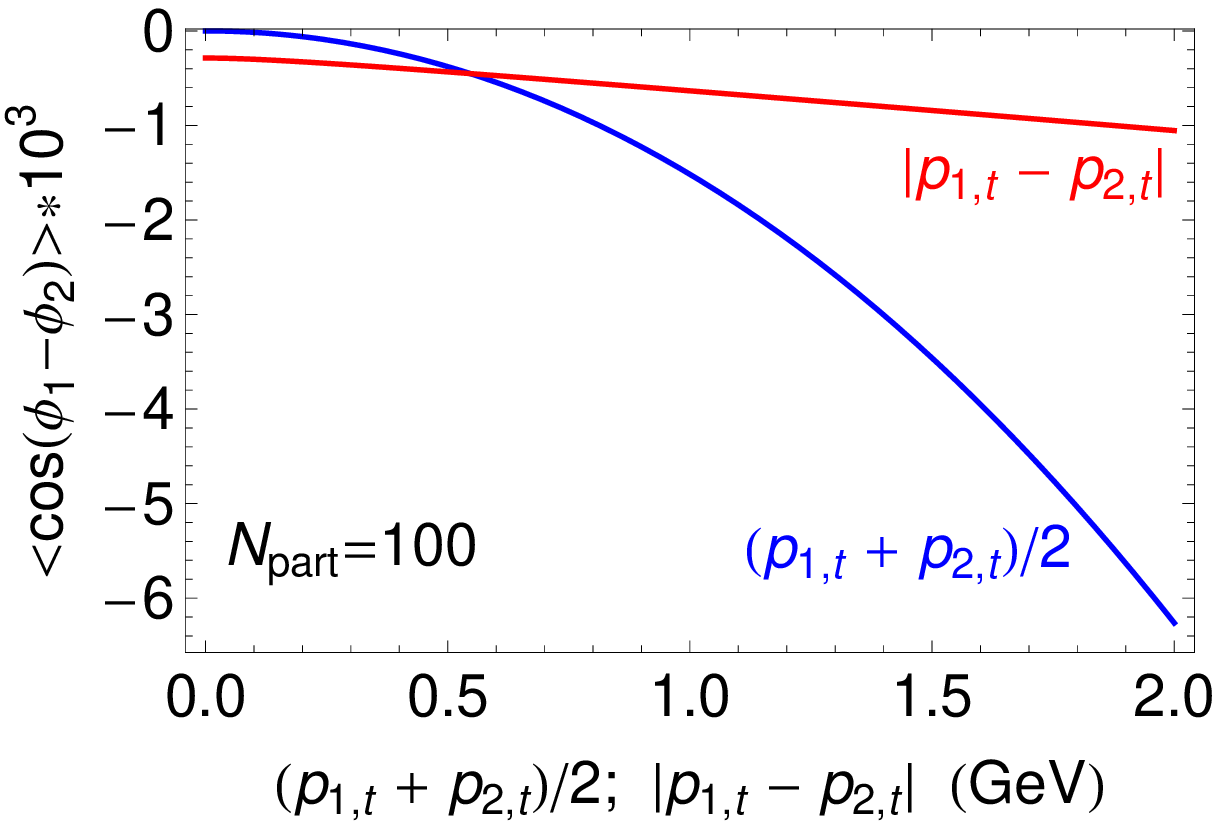}
\end{center}
\caption{(Color online) The two-particle azimuthal correlation $\left\langle
\cos (\protect\phi _{1}-\protect\phi _{2})\right\rangle $ vs $%
p_{+}=(p_{1,t}+p_{2,t})/2$ (blue line) and $p_{-}=\left|
p_{1,t}-p_{2,t}\right| $ (red line) for $N_{part}=100$.}
\label{fig2}
\end{figure}

We also performed full calculations with $g(p_{t},\eta )$, $h(\eta )$ and $%
v_{2}(p_{t},\eta )$ discussed in the previous Section and given by Eqs. (\ref%
{g}), (\ref{h}) and (\ref{v2_app}), respectively. However, in this
calculation we correct $v_{2}(p_{t},\eta )$ and assume a constant value for $%
v_2$ for transverse momenta $p_{t}>2$ GeV. The results for $N_{part}=100$
are presented in Fig. \ref{fig1} and Fig. \ref{fig2} for $\left\langle \cos
(\phi _{1}+\phi _{2})\right\rangle $ and $\left\langle \cos (\phi _{1}-\phi
_{2})\right\rangle $, respectively. 

Again, similarly to what has been observed in the STAR data, the correlation 
$\left\langle \cos (\phi _{1}+\phi _{2})\right\rangle $ grows rapidly with
increasing $p_{+}$ while appears much flatter with increasing $p_{-}$.
Comparing our results with the STAR data we see that the transverse momentum
conservation gives comparable signal for $p_{+}$ and $p_{-}$ larger than $1$
GeV and underestimate the data for lower values of $p_{+}$ and $p_{-}$.

Finally let us discus the pseudorapidity dependence of the correlation $%
\left\langle \cos (\phi _{1}+\phi _{2})\right\rangle $. The dependence of $%
\left\langle \cos (\phi _{1}+\phi _{2})\right\rangle $ on $\left| \eta
_{1}-\eta _{2}\right| $ has been measured by STAR and found to be dominated
by $\left| \eta _{1}-\eta _{2}\right| <2$, which is consistent with the CME
expectation. It is quite clear that such dependence cannot be obtained in
the present calculation, as no pseudorapidity dependence appears in the
nontrivial part of the two-particle correlation function shown in Eq. (\ref%
{f_2end}). Consequently, transverse momentum conservation predicts the
correlator $\left\langle \cos (\phi _{1}+\phi _{2})\right\rangle $ to be
essentially flat as a function of $\left| \eta _{1}-\eta _{2}\right| $ in
the midrapidity region except for a very mild dependence due to the slight
dependence of $f(\vec{p})$ and $v_{2}$ on $\eta $. However, here we have
assumed that the transverse momentum is balanced over the entire rapidity
interval. In an actual heavy ion reaction it is not unreasonable to expect
that the transverse momentum is balanced over a shorter rapidity interval.
If this were the case, we would predict not only a stronger rapidity
dependence of the signal but also a considerably stronger signal at
midrapidity. Therefore, it would be worthwhile to construct and measure an
equivalent of the charge balance function \cite{balance} for the transverse
momentum. This problem is currently under our consideration.

\section{Conclusions and comments}

We have quantitatively investigated the contribution of transverse momentum
conservation to the azimuthal correlation observables $\gamma =\left\langle
\cos (\phi _{1}+\phi _{2})\right\rangle $ and $\delta =\left\langle \cos
(\phi _{1}-\phi _{2})\right\rangle $ measured by the STAR collaboration as
motivated by the possible strong local parity violation and Chiral Magnetic
Effect. Our conclusions can be summarized as follows.

\begin{itemize}
\item[(i)] The contribution due to transverse momentum conservation is
comparable in magnitude to the prediction of the Chiral Magnetic Effect as
well as the data. In the STAR acceptance we find this contribution to be
approximately equal to $\gamma \approx -1.7v_{2}/N$, where $v_{2}$ is the
elliptic flow coefficient at midrapidity and $N$ is the total number of
produced particles (neutral and charged). This result suggests rather week
energy dependence at RHIC since $v_{2}$ and $N$ scale similarly with energy.

\item[(ii)] Taking $v_{2}=0.06$ and $N=21N_{part}$ where $N_{part}$ is the
number of participants we obtained $\gamma \cdot N_{part}\approx -0.005$,
which is a factor $3-5$ (very peripheral -- mid-central) smaller than the
experimental data. Thus we may conclude that the transverse momentum
conservation alone cannot explain the data. Also there is no
charge-dependence as opposed to experimental data. It is however a
significant source of background that eventually must be quantified and
taken into account if one really wants to extract the possible CME from the
present (and future) data.

\item[(iii)] We have demonstrated that finite acceptance issues, i.e. the
facts that particles are measured in a relatively narrow pseudorapidity bin
and neutral particles are not detected, do not suppress the effect of
transverse momentum conservation on $\gamma $.

\item[(iv)] We studied the dependence of $\gamma $ vs $%
p_{+}=(p_{1,t}+p_{2,t})/2$ and $p_{-}=\left| p_{1,t}-p_{2,t}\right| $. We
found that $\gamma $ increases with increasing $p_{+}$, $\gamma \propto (%
\frac{p_{+}}{\left\langle p_{t}\right\rangle })^{\alpha _{+}}$ with $\alpha
_{+}=2-3$. The dependence on $p_{-}$ is much weaker $\gamma \propto (\frac{%
p_{-}}{\left\langle p_{t}\right\rangle })^{\alpha _{-}}$ with $\alpha
_{-}\approx 1$. This behavior is qualitatively similar to what is observed
in the data. We also investigated the dependence of $\gamma $ on $|\eta
_{1}-\eta _{2}|$ and found no pseudorapidity dependence in contrast to what
is observed in the data.

\item[(v)] Finally we calculated $\delta =\left\langle \cos (\phi _{1}-\phi
_{2})\right\rangle $ and found that in the STAR acceptance $\delta \approx
-1/N$. We found this contribution to be a factor $1.5-4$ (mid-central --
very peripheral) \textit{larger} than the experimental data indicating again
that the transverse momentum conservation effect is a significant source of
background.

\item[(vi)] The present calculation is based on the minimal assumption that
the transverse momentum is balanced over \textit{all} particles (in the full
phase space). Thus it is very likely that the calculated contribution to $%
\gamma $ and $\delta $ from the transverse momentum conservation represents
rather the lower limit. Should, as it is not reasonable to assume, the
transverse momentum be balanced over a finite rapidity interval, we predict
not only a stronger effect at midrapidity but also a rapidity dependence of
the correlation function $\gamma $ and $\delta $. Thus a measurement of
something like a transverse momentum balance function would be highly
desirable. This problem is currently under our consideration.
\end{itemize}

We end this paper with a few further comments.

\begin{itemize}
\item[(a)] While transverse momentum conservation alone is not sufficient to
explaining the data, one may combine it with other effects such as the
Chiral Magnetic Effect or Local Charge Conservation \cite{SS-SP} to get
closer to the data. In Table. \ref{table_summary} we summarize the estimated
contributions to the azimuthal correlations from these effects together with
the STAR data. All numbers quoted are for $AuAu$ $200$ GeV collisions at
about $50-60\%$ centrality (corresponding to $N_{part}\approx 50$ \cite%
{STAR-long}). 
\begin{table}[h]
\begin{tabular}{|c|c|c|c|c|}
\hline
$\hat{O}\times 10^{3}$ & $\left\langle \cos (\phi _{1}+\phi
_{2})\right\rangle _{++}$ & $\left\langle \cos (\phi _{1}+\phi
_{2})\right\rangle _{+-}$ & $\left\langle \cos (\phi _{1}-\phi
_{2})\right\rangle _{++}$ & $\left\langle \cos (\phi _{1}-\phi
_{2})\right\rangle _{+-}$ \\ \hline
CME & $-(0.1-1)$ & $+(0.01-0.1)$ & $+(0.1-1)$ & $-(0.01-0.1)$ \\ 
LCC & $\sim 0$ & $+(0.1-1)$ & $\sim 0$ & $+(1-10)$ \\ 
TMC & $\sim -0.1$ & $\sim -0.1$ & $\sim -1$ & $\sim -1$ \\ \hline
DATA & $-0.45$ & $+0.06$ & $-0.38$ & $+1.97$ \\ \hline
\end{tabular}%
\caption{Estimated contributions to azimuthal correlations from various
effects and comparison with data. CME, LCC, and TMC stand for Chiral
Magnetic Effect, Local Charge Conservation, and Transverse Momentum
Conservation, respectively, while the DATA is from STAR measurement for $%
AuAu $ $200$ GeV collisions at about $50-60\%$ centrality.}
\label{table_summary}
\end{table}

As a precaution, all numbers (except the STAR data) bear considerable
uncertainty. The numbers from Chiral Magnetic Effect (CME) for observables $%
\left\langle \cos (\phi _{1}+\phi _{2})\right\rangle _{++,+-}$ are extracted
from \cite{cme1}. The numbers for $\left\langle \cos (\phi _{1}-\phi
_{2})\right\rangle _{++,+-}$ where obtained using the relation $\left\langle
\cos (\phi _{1}-\phi _{2})\right\rangle _{++}=-\left\langle \cos (\phi
_{1}+\phi _{2})\right\rangle _{++}$ and analogously for $(+,-)$, which hold
in case of a pure CME. The numbers from Local Charge Conservation (LCC) are
inferred from \cite{SS-SP}: the authors showed the difference $\left\langle
\cos (\phi _{1}+\phi _{2})\right\rangle _{+-}-\frac{1}{2}\left[ \left\langle
\cos (\phi _{1}+\phi _{2})\right\rangle _{++}+\left\langle \cos (\phi
_{1}+\phi _{2})\right\rangle _{--}\right] $. From LCC one should expect $%
\left\langle \cos (\phi _{1}+\phi _{2})\right\rangle _{++,--}\sim 0$ so the $%
\left\langle \cos (\phi _{1}+\phi _{2})\right\rangle _{+-}$ could be
inferred. Furthermore the value for $\left\langle \cos (\phi _{1}-\phi
_{2})\right\rangle _{+-}$ is estimated by $\left\langle \cos (\phi _{1}-\phi
_{2})\right\rangle _{+-}\sim \left\langle \cos (\phi _{1}+\phi
_{2})\right\rangle _{+-}/v_{2}$. Finally as the authors pointed out, their
results represent an upper limit of the magnitude for LCC as they enforce
exactly local charge neutrality. Our results for Transverse Momentum
Conservation (TMC) are as given in the previous sections (with the expected
uncertainty within a few tens of percent). The STAR data is from \cite%
{star-p}. Even though these are rough estimates, one still can make a few
observations: first, no single effect shows a pattern for all observables
that are in accord with data; second, different correlators appear to be
dominated by different effects, in particular -- both CME and TMC provide
important contributions to $\left\langle \cos (\phi _{1}+\phi
_{2})\right\rangle _{++}$, while TMC seems necessary for explaining the $%
\left\langle \cos (\phi _{1}-\phi _{2})\right\rangle _{++}$, and LCC for the
observed value of $\left\langle \cos (\phi _{1}-\phi _{2})\right\rangle
_{+-} $. The situation for $\left\langle \cos (\phi _{1}+\phi
_{2})\right\rangle _{+-}$ on the other is more complicated and none of the
effects discussed here seems to dominate. Clearly, additional measurements
will be required to disentangle this situation.

\item[(b)] As far as comparisons of the data with models is concerned, such
as the ones presented by the STAR collaboration \cite{star-p}, one has to
ensure that the model satisfies at least two criteria. First, the model has
to conserve transverse momentum not only on average but event-by-event.
Second, the model has to reproduce the measured magnitude of the elliptic
flow, $v_{2}$, as essentially all ``trivial'' contributions to $\left\langle
\cos (\phi _{1}+\phi _{2})\right\rangle _{++,+-}$ scale with $v_{2}$. For
example, UrQMD is known to underestimate the measured $v_{2}$ \cite{URQMD}.
This may be partly the reason that it also underestimates the measured data
for $\left\langle \cos (\phi _{1}+\phi _{2})\right\rangle _{++}$ as reported
in Ref. \cite{star-p}.
\end{itemize}

\bigskip

\newpage
\textbf{Acknowledgments}

This work was supported in part by the Director, Office of Energy Research,
Office of High Energy and Nuclear Physics, Divisions of Nuclear Physics, of
the U.S. Department of Energy under Contract No. DE-AC02-05CH11231 and by
the Polish Ministry of Science and Higher Education, grant No. N202 125437.
A.B. also acknowledges support from the Foundation for Polish Science
(KOLUMB program).

\end{document}